\documentclass[twocolumn,twoside,slac_two]{revtex4}
\usepackage{graphicx}
\usepackage{fancyhdr}
\begin{document}

\title{Muon Cooling and Future Muon Facilities: The Coming Decade}

%

\author{Daniel M. Kaplan}
\affiliation{Physics Division, Illinois Institute of Technology, Chicago, IL 60616, USA}

\begin{abstract}
Muon colliders and neutrino factories are attractive options for future facilities aimed at achieving the highest lepton-antilepton collision energies and precision measurements of parameters of the neutrino mixing matrix. The performance and cost of these depend sensitively on how well a beam of muons can be cooled. Recent progress in muon cooling design studies and prototype tests nourishes the hope that such facilities can be built in the decade to come.
\end{abstract}

\maketitle

\thispagestyle{fancy}


\section{Introduction}
While muon colliders have been discussed since the 1960s~\cite{Tikhonin-Budker,MC}, only recently has the needed technology been understood clearly enough for a concrete 
plan to be developed. Muons offer important advantages over electrons. Radiative processes are substantially suppressed, allowing acceleration and collision in rings\,---\,greatly reducing the footprint and cost\,---\,as well as a more monochromatic collision energy and potential feasibility at much higher energies.
A five-year plan for muon collider R\&D is now available~\cite{5yp} and could lead to the start of facility construction by the end of the coming decade (the 2010s).

Neutrino factories are a more recent idea~\cite{Geer}. Muons decaying in a storage ring constitute a unique source of well-characterized electron and muon neutrinos and antineutrinos, allowing  comprehensive tests of neutrino mixing~\cite{Bross,Albright}. Arguably no {\em new} technology is needed; the R\&D program is focused more  on issues of performance and cost than on feasibility {\it per se}~\cite{NFcost}. Given the decision to build one, a neutrino factory could be operational by the end of the coming decade.

\begin{table}
\caption{Representative parameters~\protect\cite{Task-Force-Report} for a low- (LEMC), medium- (MEMC), or high-emittance (HEMC) 1.5\,TeV center-of-mass-energy muon collider.
\label{tab:paramsmu}}
\begin{tabular}{|l|ccc|c|}
\hline
& LEMC & MEMC & HEMC &\\
$\cal L$ & 2.7 & 1.3--2 & 1 & $10^{34}$cm$^2$sec$^{-1}$\\
\hline
$\Delta\nu$ & 0.05 & 0.09 & 0.1  & \\
Rep. rate & 65 & 40--60 & 13 & Hz\\
$\mu$/bunch & 1 & 11  &20 &$10^{11}$\\
\# bunches & 10 & 1 &  1 & \\
Storage-ring $<$$B$$>$ & 10 & 6 & 6 & T\\
$\beta^*~(=\sigma_z)$ & 0.5 & 1 & 1 & cm\\
$\left.\!\frac{dp}{p}\right\vert_{rms}$ & 1 & 0.2 & 0.1 & \%\\
$\mu$ survival & 31 & 20 & 7 & \%\\
Colliding $\mu$/P.O.T. & 4.7 &3 &1 & \% \\
$\epsilon_{\perp,n}$ & 2.1 & 12 & 25 & $\pi$\,mm$\cdot$mrad \\
$\epsilon_{||,n}$ & 0.35 & 0.14 & 0.07 & $\pi$\,m \\
\hline
\end{tabular}
\end{table}

\begin{table}
\caption{Representative neutrino factory parameters \protect\cite{ISS}.
\label{tab:paramsnu}}
\begin{tabular}{|l|c|c|}
\hline
{$\mu^\pm$ decays/year/baseline} & $5\times10^{20}$ & \\
\hline
$P_{\rm driver}$ & 4 & MW\\
Rep. rate & 50 & Hz\\
$E_p$ & $10\pm5$ & GeV\\
\hline
Decay-channel length & 100 & m\\
Buncher length & 50 & m\\
Phase-rotator length & 50 & m\\
\hline
Cooling-channel length & 80 & m\\
RF frequency & 201.25 & MHz \\
Absorber material & LiH & \\
Absorber thickness per cell & 1 & cm \\
Input emittance  & 17 & $\pi$\,mm$\cdot$rad \\
Output emittance & 7.4 & $\pi$\,mm$\cdot$rad \\
Central momentum & 220 & MeV/$c$ \\
\hline
Final muon energy & 25 & GeV \\
\hline
Number of decay rings & 2 & \\
Decay-ring circumference & 1,609 & m\\
Straight-section length & 600 & m\\
\hline
\end{tabular}
\end{table}

\begin{figure*}[t]
\includegraphics[width=6in]{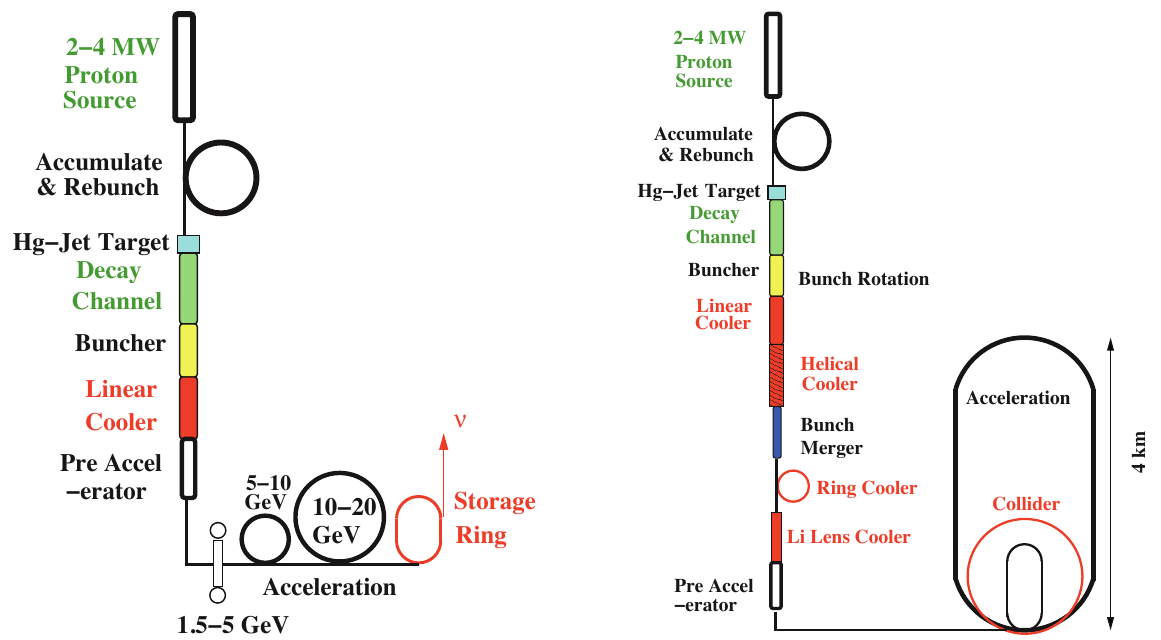}
\caption{Comparison of 20\,GeV neutrino factory (left) and 1.5\,TeV muon collider (right). The ``front end" (muon production, collection, bunching, bunch rotation, and initial cooling) can be the same for both. It is followed in a neutrino factory by acceleration of the muons to multi-GeV energy and injection into a storage ring, with long straight sections in which muon decay forms intense neutrino beams aimed at near and far detectors. For a muon collider, the front end is followed by 6-dimensional cooling, bunch coalescence, and acceleration to high energy (e.g., 0.75\,TeV) for injection into a collider ring, where $\mu^+$ and $\mu^-$ bunches collide for $\sim10^3$ turns.}\label{fig:facilities}
\end{figure*}

The two types of facility are schematically compared in Fig.~\ref{fig:facilities} and are seen to have much in common. Representative parameters are listed in Tables~\ref{tab:paramsmu} and \ref{tab:paramsnu}.
For both types, the performance and cost depend sensitively on how well a muon beam  can be cooled. 

Neutrino factories might be feasible without cooling~\cite{nocool}, but  transverse cooling of the muon beam by up to about an order of magnitude in six-dimensional (6D) beam emittance has been shown to be cost-effective. Muon colliders require much more substantial cooling\,---\,a factor 10$^6$ or so in 6D emittance\,---\,in order to achieve the $\sim$\,10$^{34}$ luminosities required for the envisaged energy-frontier physics program. Collider designs at center-of-mass energies of 1.5, 4, 8\,TeV~\cite{collider-scheme} and beyond~\cite{King} have been considered. (At the highest energies  neutrino-induced off-site radiation becomes a concern; although there are strategies to mitigate this, the problem is not yet an immediate one and its solution has not been studied in detail.) Attention has also been given to less ambitious machines, e.g., a Higgs factory~\cite{MC}, or a $Z$~\cite{Marciano} or $Z^\prime$ factory~\cite{Eichten}, which could profitably operate at lower  luminosity (as well as energy), thus possibly with less cooling as well.



\section{Muon Cooling}
The short lifetime of the muon (2.2\,$\mu$s at rest) vitiates all  beam-cooling methods  currently in use (electron, stochastic, and laser cooling). However, a method almost uniquely applicable to the muon\,---\,ionization cooling~\cite{cooling}\,---\,appears adequate to the challenge. In this, muons are made to pass through material of low atomic number in a suitable focusing magnetic field; the normalized transverse emittance $\epsilon_{\perp,n}$ then obeys~\cite{Neuffer-Yellow}
\begin{equation}
\frac{d\epsilon_{\perp,n}}{ds}\ \approx\
-\frac{1}{\beta^2} \frac{dE_{\mu}}{ds}\ \frac{\epsilon_{\perp,n}}{E_{\mu}}\ +
\ \frac{1}{\beta^3} \frac{\beta_{\perp} (0.014\,{\rm GeV})^2}{2\ E_{\mu}m_{\mu}\ L_R}\,,
\label{eq1}
\end{equation}
where  $\beta = v/c$ is the muon velocity,
$\beta_{\perp}$ the
betatron function (focal length)
at the absorber, $dE_{\mu}/ds$ the energy loss per unit length, 
 $m_\mu$ the muon mass, 
and $L_R$
the radiation length of the absorber material.  (This is the expression appropriate to the cylindrically symmetric case  of solenoidal focusing, for which $\beta_x=\beta_y\equiv\beta_\perp$ and cooling occurs 
equally in the $x$-$x^\prime$ and $y$-$y^\prime$ phase planes.) The first term in Eq.~\ref{eq1} is
the cooling term, and the second is the heating term due to multiple
scattering. The heating term is minimized via small $\beta_{\perp}$ 
(strong focusing) and large $L_R$  (low-$Z$ absorber material).
For a given cooling-channel design, equilibrium emittance is  achieved when the heating and cooling terms
balance.

\begin{figure}[h]
\centering
\includegraphics[width=82mm, bb=20 80 755 458,clip]{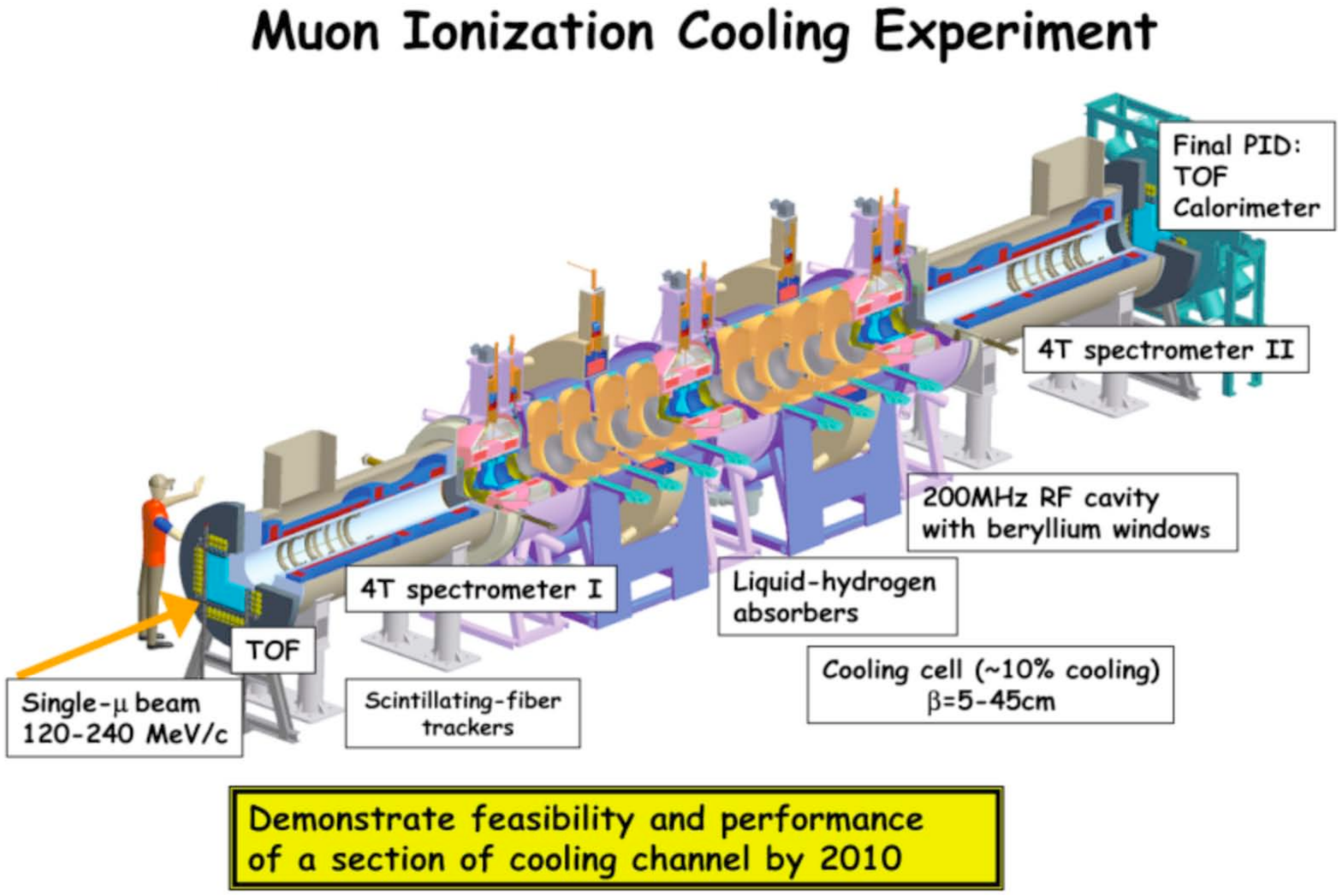}
\caption{Cutaway rendering of the Muon Ionization Cooling Experiment (MICE): one lattice cell of a cooling channel is shown, employing two liquid-hydrogen (LH$_2$) absorbers (dark blue) interspersed with radio-frequency (RF) cavities (orange), with a third absorber added at the end for symmetry and to shield the scintillating-fiber tracking detectors in the solenoidal spectrometers from RF-cavity x-ray emissions. The beam is focused to low beta at the absorbers by several superconducting coils (red). MICE will demonstrate about 10\% transverse emittance reduction of a muon beam.} \label{fig:MICE}
\end{figure}

\subsection{Muon Ionization Cooling Experiment}

Figure~\ref{fig:MICE} shows one cell of a typical ionization-cooling lattice\,---\,that of the Muon Ionization Cooling Experiment~\cite{MICE} (MICE)\,---\,surrounded by the input and output spectrometers and particle-identification detectors that will be used to demonstrate and characterize the ionization-cooling process experimentally at Rutherford Appleton Laboratory in the UK. MICE is designed to test transverse-emittance cooling of a low-intensity muon beam by measuring each muon individually. 
It will thereby demonstrate that the process is well understood in both its physics and engineering aspects, and works as simulated. The full results from MICE are expected by about 2013, with analyses of some configurations available one to two years earlier.

\section{6-Dimensional Cooling}

\begin{figure}
\includegraphics[width=70mm]{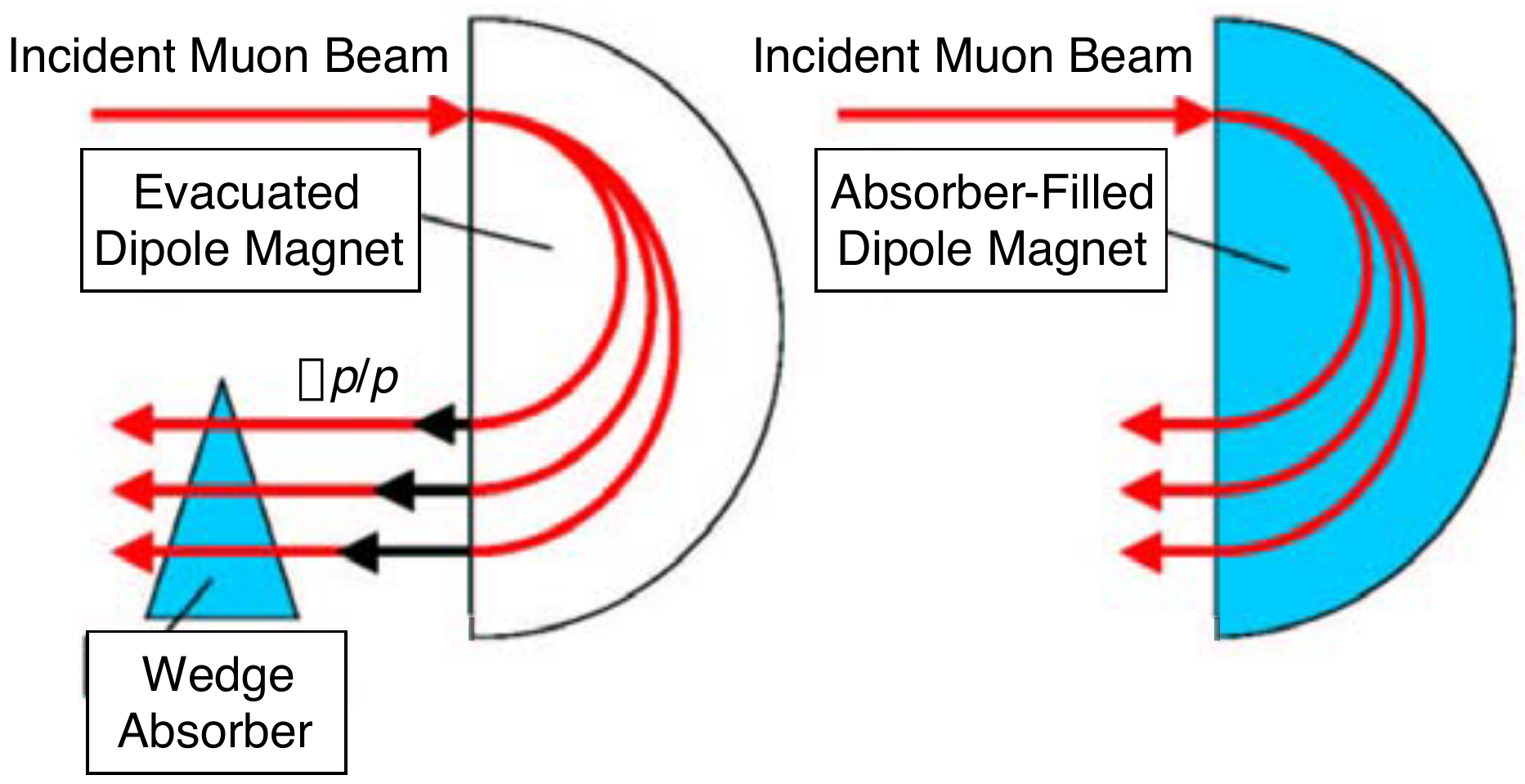}
\caption{Two approaches to emittance exchange: in each, an initially small beam with nonzero momentum spread is converted into a more monoenergetic beam with a spread in transverse position. (Figure courtesy of Muons, Inc.)}\label{fig:emittExch}
\end{figure}

\begin{figure*}[t]
\includegraphics[width=35mm]{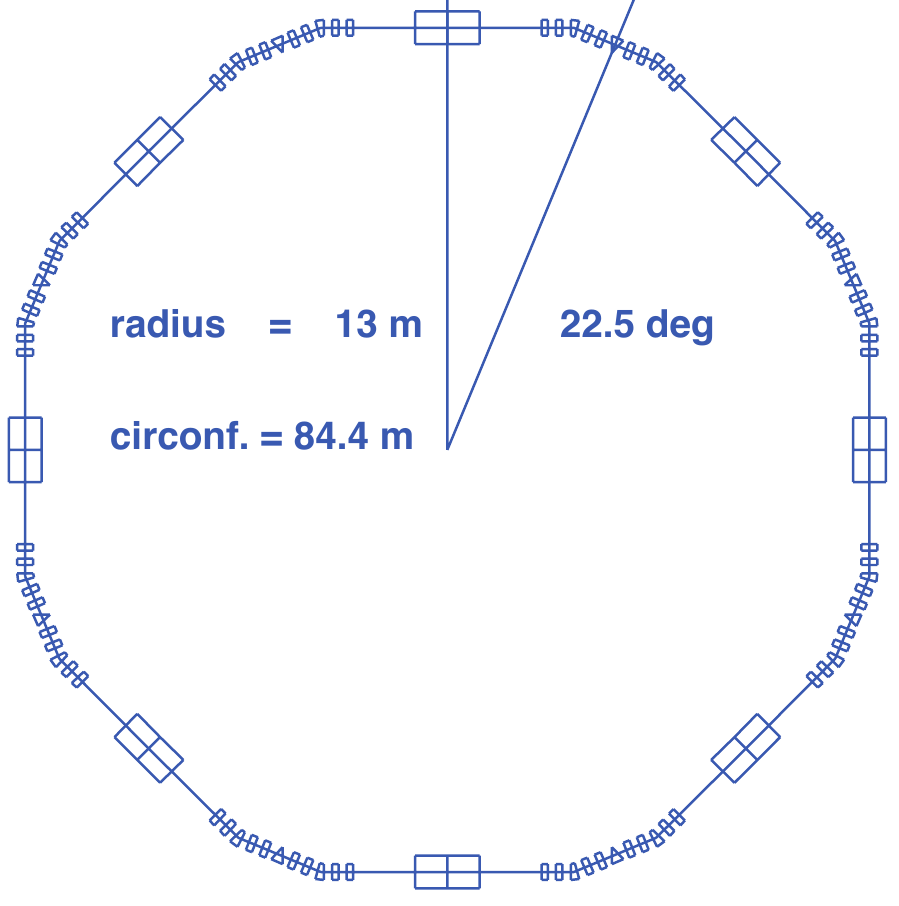}
\includegraphics[width=55mm]{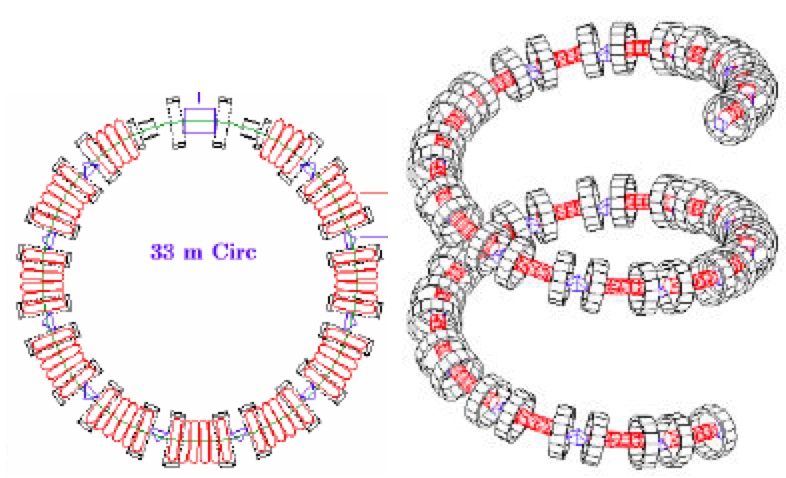}
\includegraphics[width=27mm]{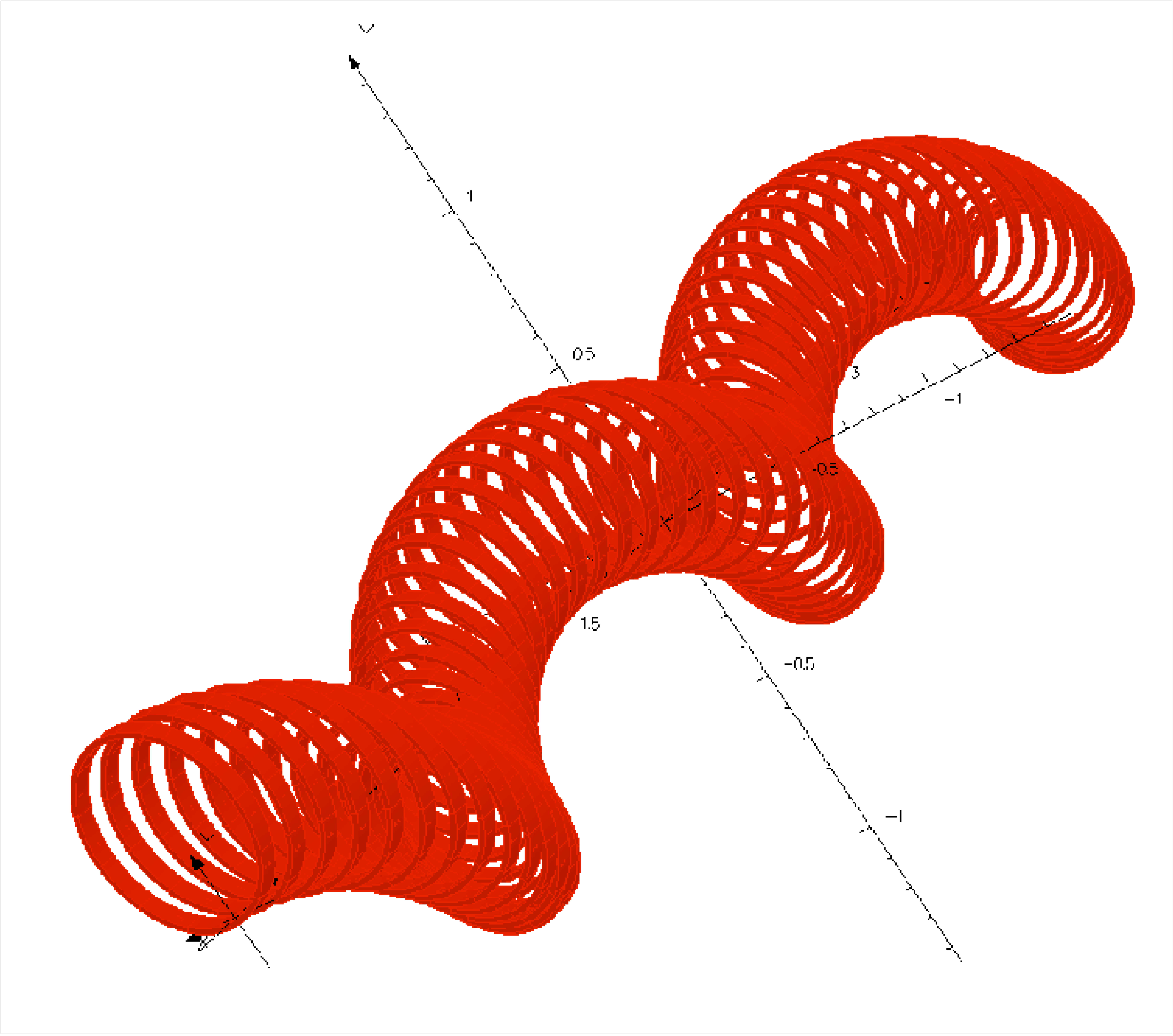}
\includegraphics[width=50mm]{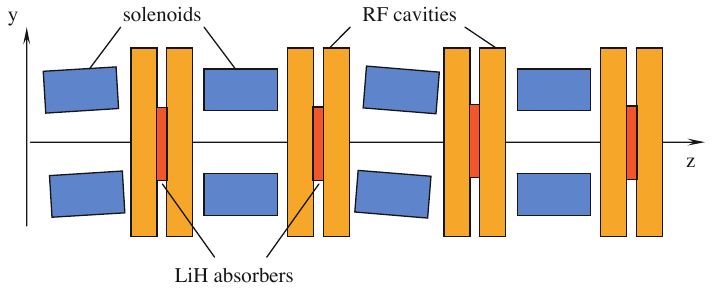}
\caption{Examples of 6D cooling apparatus that have been shown to work in simulation: (left to right) quadrupole--dipole ring, ``RFOFO" solenoid-focused ring, ``RFOFO Guggenheim" helix, helical-solenoid channel, ``snake" channel.}\label{fig:6Dcool}
\end{figure*}

As already mentioned, a high-luminosity (${\cal L}\stackrel{>}{_\sim}10^{34}$\,cm$^{-2}$s$^{-1}$) muon collider requires a more ambitious cooling scheme, reducing both transverse and longitudinal emittances by an overall factor of at least $10^6$ in 6D emittance. Several approaches to achieving this goal have been developed~\cite{collider-scheme,MuInc-collider} by the  Neutrino Factory and Muon Collider Collaboration (NFMCC) working in concert with the Fermilab Muon Collider Task Force (MCTF) and two small R\&D firms with SBIR/STTR~\cite{SBIR} funding: Muons, Inc.~\cite{MuonsInc} and Particle-Beam Lasers~\cite{PBL}. Since ionization cooling is normally effective in only the transverse phase planes, 6D emittance reduction is typically accomplished via transverse--longitudinal emittance exchange: dispersion is used to create a correlation between path length in an energy-absorbing medium and momentum (Fig.~\ref{fig:emittExch}), reducing beam energy spread at the expense of transverse emittance growth. Three general approaches have been shown to work in simulation: rings, helices, and snakes  (Fig.~\ref{fig:6Dcool}). Like transverse cooling lattices, most 6D-cooler designs employ superconducting-solenoid focusing and benefit from the ability of such  solenoids to accommodate a large aperture, generate low $\beta$, and focus simultaneously in both $x$ and $y$, enabling compactness that minimizes muon decay in flight.

The earliest successful example of a 6D cooling channel was the 4-sided solenoid-focused ring of Balbekov~\cite{tetra-ring,Alsharoa}, but it was so tightly packed as to lack space for beam injection and extraction. This first ``in-principle" success led to the development of rings with space allocated for these functions~\cite{Palmer-ring,Cline-Garren}, and to helices~\cite{HCC,Guggenheim}, which can embody the symmetries of rings, but are open at the ends for muon ingress and egress and reduce beam loading on absorbers and RF cavities. Helices can also provide faster cooling by allowing the focusing strength to increase along the channel, decreasing the equilibrium emittance as the beam is cooled. The Helical Cooling Channel (HCC), based on a Hamiltonian theory~\cite{HCC}, uses a combination of ``Siberian Snake" helical dipole and solenoid fields and 
employs a continuous, high-pressure, gaseous-hydrogen absorber so as to minimize both the deleterious effects of windows and (via pressurized RF cavities, discussed below) the length of the channel.  Subsequent to the invention of the HCC, it was shown that its required solenoid, helical dipole, and (for increased acceptance) helical quadrupole  field components can be produced by a simple sequence of offset current rings
~\cite{Kashikhin} (Fig.~\ref{fig:6Dcool}, right). The ``Snake" channel~\cite{Alexahin} (Fig.~\ref{fig:6Dcool}, far right) is the ``least circular" of these approaches and brings the economy of simultaneously accommodating muons of both signs.

\begin{figure}
\includegraphics[width=2in]{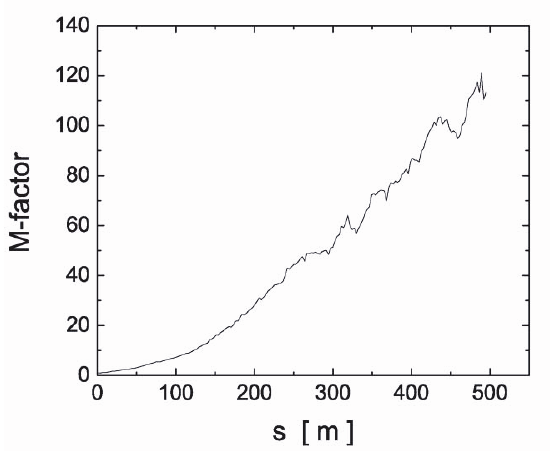}
\caption{Merit factor (defined in text) vs.\ path length for idealized RFOFO cooling ring of \protect\cite{Palmer-ring}.}\label{fig:merit}
\end{figure}

To compare a given proposed muon-cooling technique with others calls for a suitable figure of merit, and different merit factors may be appropriate depending on the details of the facility and the physics at which it is aimed. One popular merit factor is 
\begin{equation}
M(s) = \frac{\epsilon_{6,n}(s)}{\epsilon_{6,n}(0)}\frac{N(s)}{N(0)}\,,
\end{equation}
where $\epsilon_{6,n}$ is the normalized 6D emittance and $N$ the total number of surviving muons, as a function of path-length $s$. This peaks at $\approx$\,120 (Fig.~\ref{fig:merit}) for the ``ideal" RFOFO cooling ring of \cite{Palmer-ring} (2nd from left in Fig.~\ref{fig:6Dcool}); for comparison, a similar calculation for an idealized MICE-like, linear, transverse-only  cooling lattice plateaus at $M\approx15$. These  merit factors reach a plateau as the beam emittance approaches the cooling channel's equilibrium emittance, and then fall off with increasing path-length as muon decay continues to reduce the beam intensity.

The value $M\approx120$ for the RFOFO ring is somewhat illusory as there were no windows in that simulation, nor any space left for beam injection and extraction\,---\, the gap shown at the top of the ring in Fig.~\ref{fig:6Dcool} was filled with a 12th cooling cell. If an injection/extraction gap is made and realistic windows put in for both the LH$_2$ absorbers and the RF cavities (see below), the merit factor falls to $\approx$\,15~\cite{Palmer-ring}. 

The schemes described above all work near the ionization minimum ($\gamma\beta\approx2$). An entirely different approach seeks to exploit the much higher ionization energy-loss rate at the ``Bragg peak" ($\gamma\beta\approx0.01$)~\cite{PDG} but has significant challenges to overcome (e.g., sufficiently rapid acceleration, and making windows thin enough to overcome multiple scattering) due to these low velocities. This ``frictional cooling" regime has been studied experimentally~\cite{Muhlbauer} and R\&D continues~\cite{frictional}. A recent conceptual advance, the ``particle refrigerator," seeks to increase the energy acceptance of the frictional cooling channel by two to three  orders of magnitude~\cite{fridge} and could lead to very compact high-flux muon sources; the technique may also be applicable to decelerating and cooling other particle species besides muons~\cite{other-species}. In contrast to the schemes discussed previously, by taking advantage of the positive slope of the $dE/dx$ curve just below the Bragg peak, frictional cooling can cool directly in 6D, with no emittance exchange necessary.

\begin{figure}
\vspace{-.075in}
\centerline{\hspace{-.1in}\includegraphics[width=3in
]{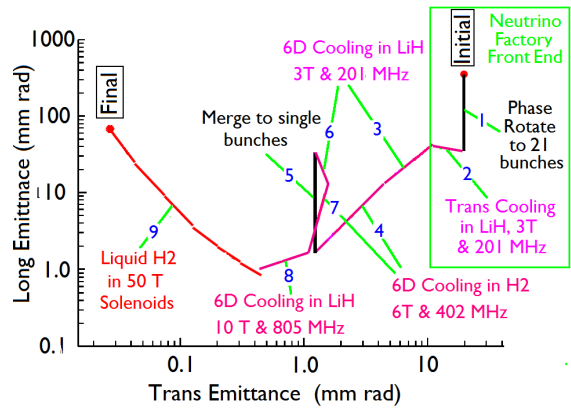}}
\vspace{-.075in}
\caption{Cooling ``trajectory" in longitudinal and transverse emittance for a particular scheme (from \protect\cite{collider-scheme}).}\label{fig:Fernow-Neuffer-plot}
\end{figure}

Various schemes can be evaluated and compared by displaying the ``cooling trajectory" on a plot of longitudinal vs.\ transverse emittance. Figure~\ref{fig:Fernow-Neuffer-plot} is an example~\cite{collider-scheme} in which step 2 is a simplified MICE-like transverse cooling lattice employing solid-LiH absorbers~\cite{NFcost} and steps 3--8 are 6D ``RFOFO Guggenheim" helices~\cite{Guggenheim}, at the end of which the muon bunches are shorter than necessary for the luminosity goal but more transverse cooling is still needed. In principle this ``final cooling" can be achieved using extremely high-field ($\stackrel{<}{_\sim}$\,50\,T) solenoids enclosing LH$_2$ absorbers, in which transverse cooling can be carried out at the expense of longitudinal emittance as the muon momentum is allowed to fall towards the Bragg peak. Although such solenoids seem feasible using high-temperature superconductor (e.g.\ Bi-2223 tape) operated at LHe temperature~\cite{HTS-solenoid}, given the large magnetic forces involved, considerable R\&D will be required in order to realize them~\cite{HTS-insert}. (Another motivation for high-field magnets for muon colliders is that the luminosity in the collider ring increases with the frequency of collisions, with stronger dipole fields giving smaller ring circumference and  more collisions per muon lifetime.) Other schemes for reaching these small transverse emittances (or yet smaller ones) have also been discussed~\cite{MuInc-collider,extreme-cool,optical}. Smaller transverse emittance can potentially give higher collider luminosity with fewer muons, thus allowing a lower-power proton driver and reducing neutrino-induced radiation as well as decay-electron background in the collider detector. The goal of a low-emittance muon collider has been substantially advanced by the recent series of workshops organized by Muons, Inc.~\cite{LEMC}.

Which combination of these approaches to cooling for a muon collider will in the end be chosen as optimal remains to be seen; ``down-selection" among alternatives would appear to be premature at present and is one of the tasks foreseen in the 5-year plan.

\subsection{6D Cooling Experiments}

It is desirable to test 6D muon cooling experimentally. A proposal to do so (MANX) has been developed~\cite{MANX}, based on the helical cooling channel, using the MICE muon beam and detectors (or possibly a new beam that could be built at Fermilab). 
An important aspect of  MANX is its applicability not only to muon colliders but also to upgrading the sensitivity of the proposed muon-to-electron conversion experiment 
at Fermilab~\cite{Mu2e}. Thus,  it may be worth carrying out in that context, independent of the muon collider R\&D plan.
Other ideas have also been discussed, ranging from operation of one or more wedge absorbers in MICE (with muons selected and weighted off-line to create a suitable momentum--position correlation for emittance exchange~\cite{Rogers-Snopok,5yp}) to constructing and testing a small-scale cooling ring~\cite{Summers-ring} or a portion of a Guggenheim or final-transverse-cooling lattice~\cite{5yp}. 

A goal of the 5-year plan~\cite{5yp} is to clarify which of the various 6D cooling approaches need to be demonstrated experimentally; a proposal for a 6D demonstration experiment is one of the plan's deliverables. Since such experiments require a substantial  investment of effort and resources, only the minimum necessary  number  should be undertaken. If MICE with wedge-absorber tests plus 6D-cooling simulation and design studies can be shown to create sufficient confidence that 6D cooling is understood, it may even be preferable to proceed directly to a muon collider design-and-construction project, with any 6D-cooling tests done as part of that project, rather than as a separate,  prerequisite effort. Both the risks and the benefits of proceeding with or without each potential experiment will need to be carefully evaluated.

\section{Other R\&D Issues}

Although muon cooling is the least familiar aspect of muon facilities, a few other issues are also prominent in the R\&D program.

\subsection{Proton Driver}

These proposed facilities require intense pulses of medium-energy protons in order to make sufficient pions for the $10^{21}$ muons/year goal. A number of designs seem capable of meeting the specification~\cite{ISS}. Generally they entail proton-beam power in the ballpark of 4\,MW\,---\,over a reasonable range in proton-beam kinetic energy (roughly 2\,--20\,GeV), production of pions (of the few-hundred-MeV energies which efficiently yield ionization-coolable muons) is approximately a function of beam power-on-target only~\cite{Strait,ISS}.

\subsection{Targetry}

Using medium-energy protons to produce so many muons requires a target system that goes well beyond the capabilities of those currently operating at the world's accelerator laboratories~\cite{Hylen}. A 4\,MW beam impinging on a solid target is likely to damage it substantially in a shorter-than-desirable time. Anything less than a several-month target lifetime will lead to undesirable operating overhead due to the multi-week delay involved in changing out a highly radioactive target surrounded by highly radioactive shielding.

\begin{figure*}
\includegraphics{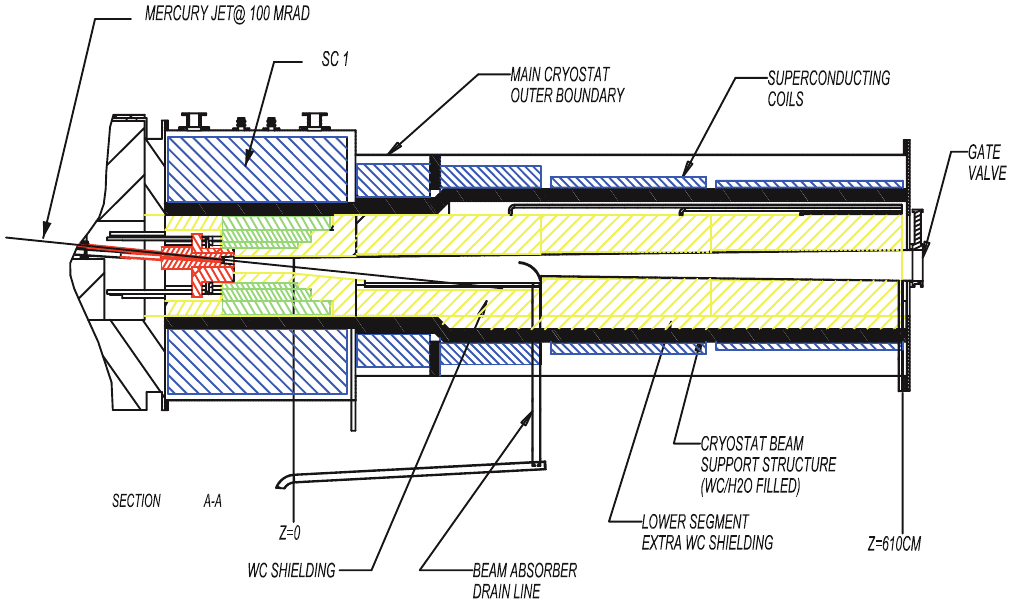}
\caption{Engineering drawing of 4\,MW Hg-jet target system for a neutrino factory, featuring Hg-pool beam dump and solenoid field tapering from 20\,T at left to $\approx3$\,T at right (from Neutrino Factory Feasibility Study-II~\protect\cite{FS2}).}\label{fig:Targetry}
\end{figure*}

A solution to this challenge has recently been demonstrated~\cite{McDonald-PAC09} in the MERIT experiment~\cite{MERIT} at the CERN nTOF facility. The solution is a free mercury jet flowing through vacuum within a high-field ($\approx20$\,T) solenoid (Fig.~\ref{fig:Targetry}). The intense proton pulse initiates a hadron shower which heats the target and disrupts it via cavitation, but the disruption occurs at a time determined by the speed of sound in mercury, long after the produced pions have escaped into the decay channel. The pulse structure of the proton beam can easily be arranged to have a sufficient gap for a new, pristine section of jet to form before the arrival of the next proton bunch. Preliminary MERIT results demonstrate power-handling capability well in excess of the 4\,MW specification~\cite{McDonald-PAC09}.

It has also been suggested that solid targets (now in use in high-power beams at  CERN, Fermilab, ISIS,  J-PARC, and PSI) may continue to be feasible up to $\approx$\,2--4\,MW power~\cite{ISS,Hylen}. For  example, design studies are in progress for a graphite neutrino production target for the Fermilab NO$\nu$A experiment with 2.3\,MW proton-beam power~\cite{Hylen}.

\subsection{Rapid Muon Acceleration}

\begin{figure}
\centerline{\hspace{.1in}\includegraphics[width=3.3in]{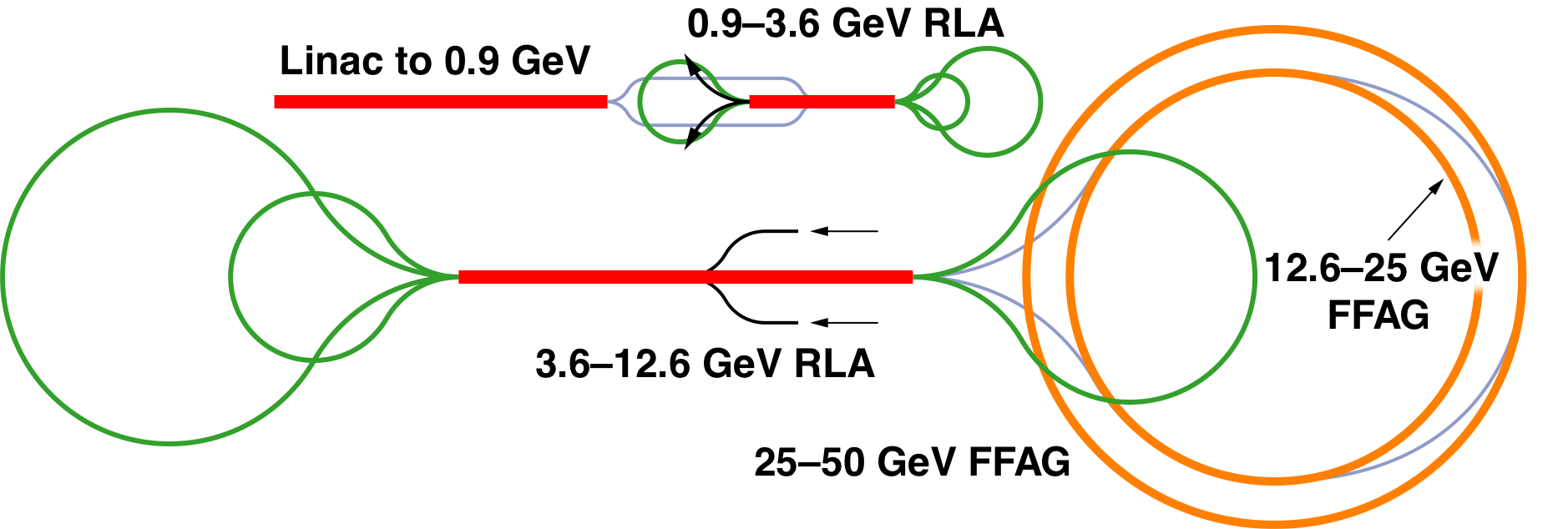
}}
\caption{Neutrino factory acceleration scheme from International Scoping Study~\protect\cite{ISS} design (including optional 25--50\,GeV acceleration stage).}\label{fig:ISS-accel}
\end{figure}

Once the muons are relativistic, time-dilation substantially suppresses decay losses. The key is then to carry out the first stages of acceleration as rapidly as possible. The proposed scheme (Fig.~\ref{fig:ISS-accel})~\cite{ISS} features a superconducting linac feeding a pair of ``dogbone" recirculating linacs feeding a non-scaling fixed-field alternating-gradient (FFAG) accelerator, bringing the muon energy to 25\,GeV. For the ultimate muon collider energies alternatives such as a very rapid-cycling synchrotron have also been considered~\cite{VRCS}. Clearly, simpler approaches (e.g., a single linac from ionization-cooling to final energy) are also feasible but would be considerably more costly.

The non-scaling FFAG is a recent innovation with novel beam-physics aspects including rapid  resonance crossing  and quasi-isochronous acceleration between RF buckets. The non-scaling feature allows small magnet apertures with concomitant cost savings. A demonstration experiment, the Electron Model with Many Applications (EMMA), is under construction at Daresbury Laboratory in the UK~\cite{EMMA}.

Scaling FFAGs for muon acceleration were proposed earlier~\cite{nocool} but featured large apertures, requiring low-gradient, low-frequency ($\sim$\,10\,MHz) RF cavities; this work inspired the non-scaling ideas. Recent progress on scaling FFAGs may lead to long, dispersion-suppressed straight sections compatible with higher-frequency RF~\cite{Mori-Planche}.

\subsection{RF Technology}

\begin{figure}
\includegraphics[width=2in]{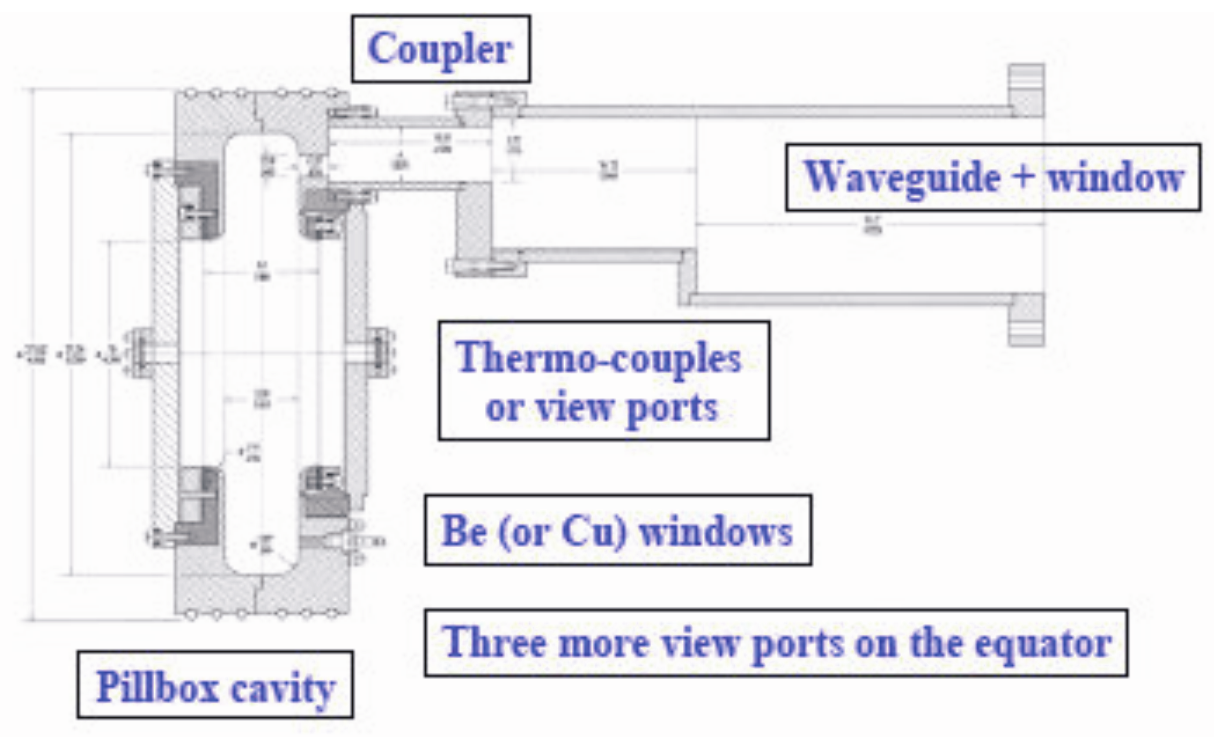}~~\includegraphics[width=1in,bb=5 -25 150 118,clip]{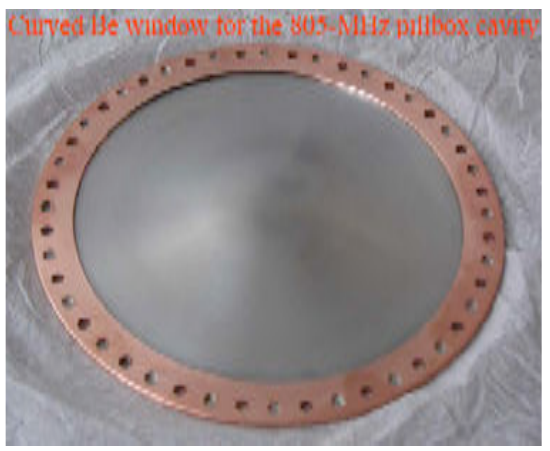}
\caption{(left) Engineering drawing of MuCool closed-cell 805\,MHz accelerating cavity; (right) photo of curved Be window for 805\,MHz cavity.}\label{fig:805}
\end{figure}

A ``cost driver" for such facilities is radio-frequency (RF) acceleration. Ionization-cooling channels require operation of RF cavities  in multi-tesla fields (precluding the use of superconducting cavities), which the ``MuCool" R\&D program has shown to be challenging~\cite{Huang,Bross}. In order to accommodate the large initial beam sizes, typical cavity frequencies are in the ballpark of 200\,MHz; however, much of the R\&D is done on ``1/4-scale" (805\,MHz) prototypes.
 These are not only easier to fabricate, test, and modify, but are also similar to those that would be used in the later stages of the cooling system, where the beam is smaller. Cavity electrical efficiency is maximized by ``pillbox" geometry, with apertures closed by thin beryllium windows (Fig.~\ref{fig:805})\,---\,a technique usable only with muons. For a given input power or maximum surface electric field, pillbox cavities have twice the accelerating gradient of standard, open-cell cavities. 
 
\begin{figure}
\centerline{\includegraphics[width=3.25in]{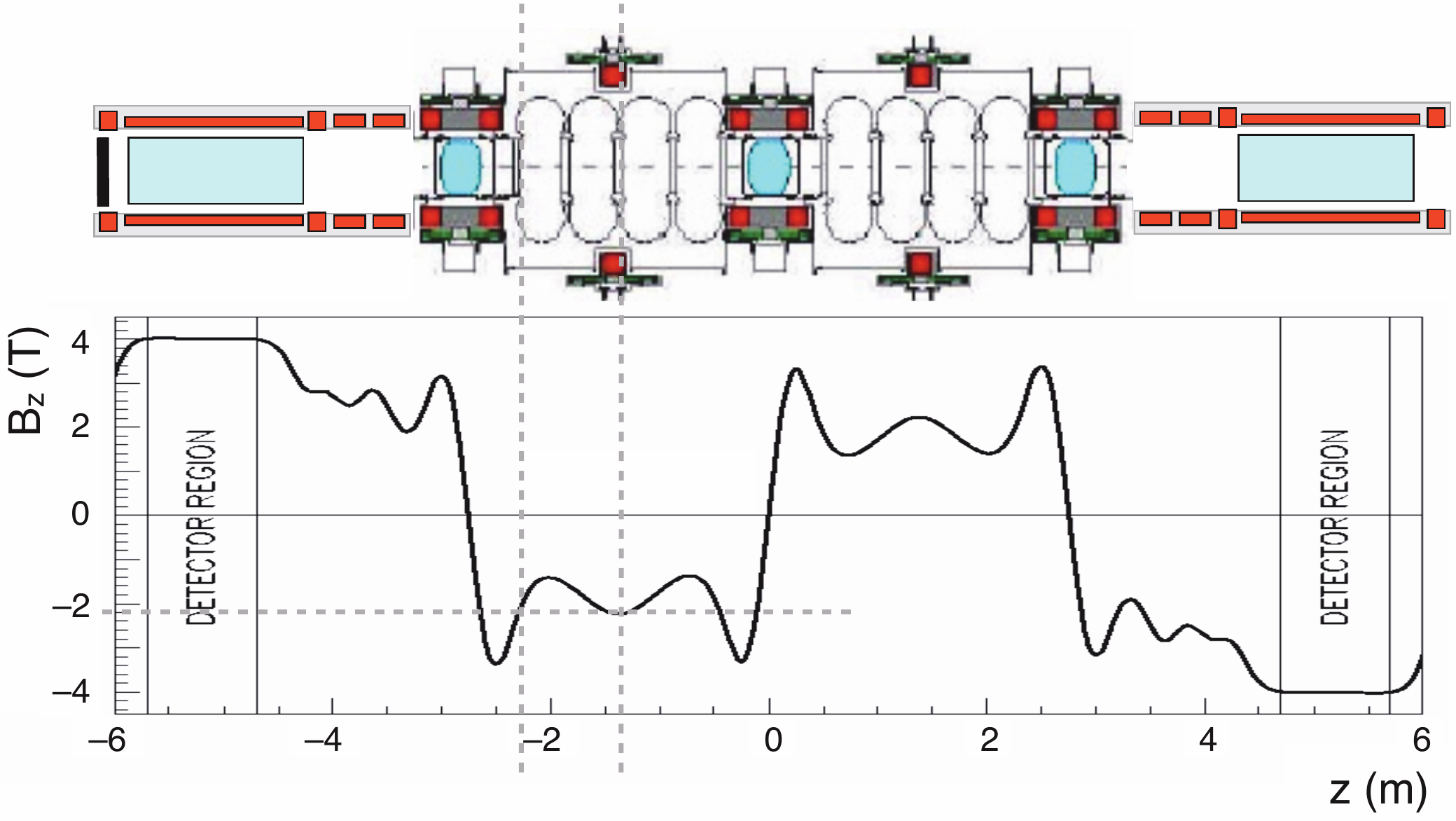}}
\caption{The maximum magnetic field at the RF-cavity windows in the MICE cooling cell is approximately 2\,T.}\label{fig:MICE-field}
\end{figure}

\begin{figure}
\vspace{.1in}
\centerline{\hspace{-.1in}\includegraphics[width=1.7in]{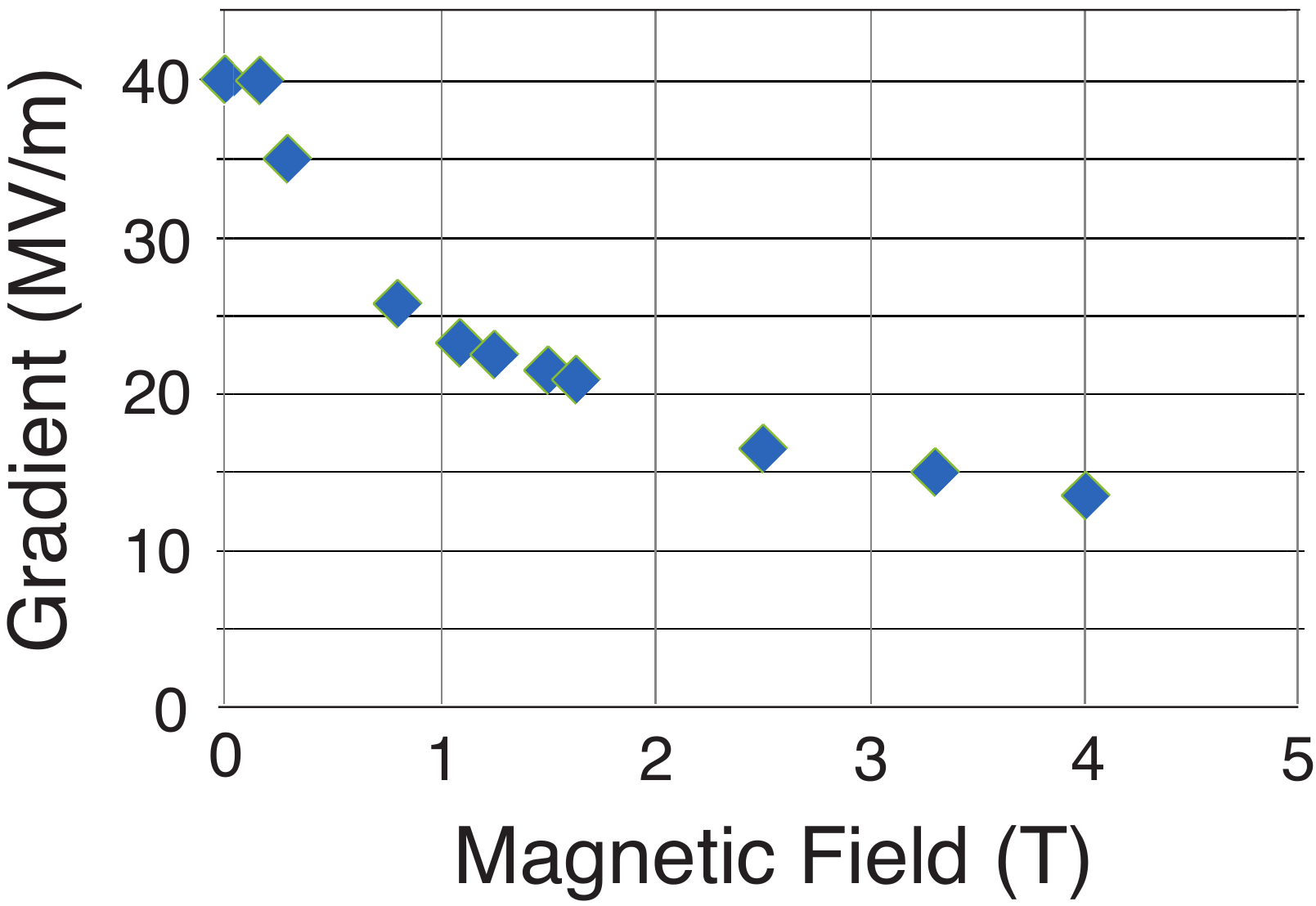}}
\caption{Observed dependence of maximum safe surface electric field (which for pillbox cavities approximately equals the on-axis accelerating gradient) vs.\ axial magnetic field for 805\,MHz copper cavities.}\label{fig:RF}
\end{figure}

To set the scale, Fig.~\ref{fig:MICE-field} shows that the maximum magnetic field on the RF-cavity windows in the MICE cooling lattice is about 2\,T; however, in later cooling stages, where lower equilibrium emittance, and hence stronger focusing, is required, the fields will need to be many times stronger.
Figure~\ref{fig:RF} shows data obtained by the MuCool R\&D collaboration\,---\,a subset of the NFMCC that is developing and testing hardware components needed for muon cooling\,---\,on an 805\,MHz copper cavity operated in a solenoidal magnetic field~\cite{Huang}. Beyond a limiting accelerating gradient, damaging sparks occur and degrade the conditioning of the cavity. The observed loss in accelerating gradient ranges from a factor of about 2 at 2\,T to 3 at 4\,T. This is not necessarily a ``show-stopper" for muon cooling but, by requiring a stretching out of the cooling channel, could impose a significant performance loss or cost increase. Techniques are being explored to mitigate the degradation, including cavity surface coatings (e.g., via Atomic Layer Deposition~\cite{ALD}), alternative cavity materials (e.g., beryllium, or beryllium-coated, cavities), open-cell cavities, and pressurized cavities. There is some indication that the degradation is related to magnetic focusing of field-emitted electrons from the window surfaces, based on early data taken with an open-cell cavity.

\begin{figure}
\includegraphics[width=3.2in]{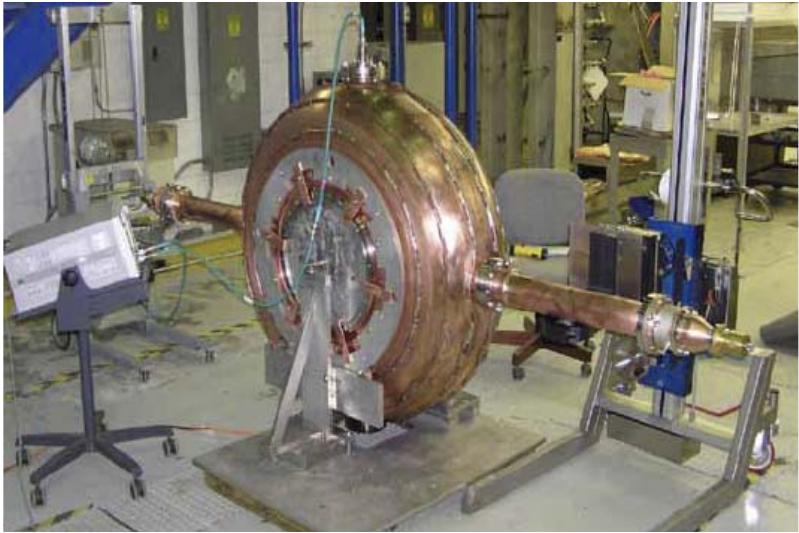}
\caption{Prototype 201-MHz copper cavity for muon cooling.}\label{fig:201}
\end{figure}

Studying the behavior of 201\,MHz cavities in magnetic field is also important, as the frequency dependence of the degradation is not known. A 201\,MHz cavity has been built (Fig.~\ref{fig:201}), and a large superconducting coil is under construction, with delivery anticipated in 2010. In the mean time data have been taken in the fringe field of the smaller magnet used for the 805\,MHz cavity tests. At up to $\approx$\,0.4\,T on the window nearest the magnet, a 25\% degradation in maximum safe gradient is observed.

\begin{figure}
\centerline{\hspace{-.1in}
\includegraphics[width=3.3in]{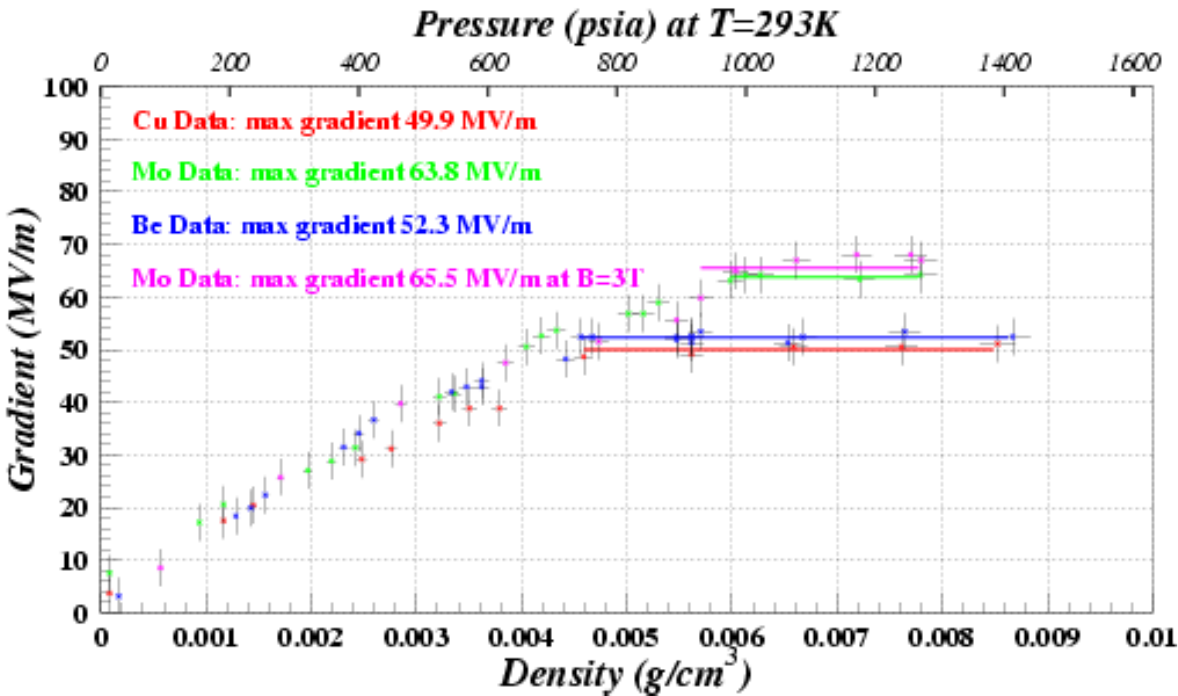}}
\caption{Observed dependence of maximum safe surface electric field in GH$_2$-pressurized 805\,MHz cavity vs.\ hydrogen density and pressure for various electrode materials. Molybdenum electrodes were tested without (green points) and with  (magenta) a 3\,T axial field, with no observed degradation in maximum safe field.}\label{fig:GH2}
\end{figure}

Cavities pressurized with hydrogen gas were initially proposed as a means of raising operating gradients via the Paschen effect~\cite{GH2}. They were subsequently found to mitigate magnetic-field-induced gradient degradation as well (Fig.~\ref{fig:GH2}). Used aggressively, they enable continuous, ``combined-function" cooling channels in which the ionization energy loss and re-acceleration take place simultaneously throughout the length of the channel~\cite{GH2,HCC}. A less ambitious application has also been suggested: using them in a ``conventional" cooling channel (e.g., that of Figs.~\ref{fig:MICE} and \ref{fig:MICE-field}) with just enough hydrogen pressure to overcome the magnetic-field-induced degradation~\cite{Zisman}. In such cavities a potential pitfall is cavity loading due to acceleration of ionization electrons~\cite{Alvin}; first experimental studies suggest that this can be overcome via a small (0.01\%) admixture of electronegative gas~\cite{Yonehara}.

Rapid muon acceleration also requires high accelerating gradient, most economically achieved by means of superconducting cavities. Large, low-frequency superconducting cavities are most economically fabricated of niobium-coated copper. However, such cavities display a ``Q disease": their resonance quality factor (and hence, electrical efficiency) degrades with increasing gradient. R\&D on this problem has been carried out at Cornell~\cite{Cornell}, with the goal of achieving $\approx$\,20\,MV/m at 201\,MHz.

\section{Outlook}

The neutrino factory is by now well studied, with two feasibility studies~\cite{FS1,FS2} and the International Scoping Study~\cite{ISS} completed and the International Design Study~\cite{IDS} (IDS) in progress. The IDS is aimed at completion by 2013, with production of a Reference Design Report, based on which an interested country or region could then commence a construction project. A key tactical question	is the size of the neutrino mixing angle $\theta_{13}$: if it is as large as a few degrees, its measurement in the Double Chooz or Daya Bay experiment could stimulate a decision to put off building a neutrino factory while multiple rounds of ``superbeam" experiments are executed. Many believe, however, that the physics of neutrino mixing will ultimately demand the unique ``resolving power" of the neutrino factory, and that particle physics will be better off if one is built sooner rather than later~\cite{Bross}.

The NFMCC/MCTF  Muon Accelerator Five-Year Plan, if funded, will produce a muon collider Design and Feasibility Study Report\,---\,the first detailed feasibility study for a muon collider. This should be followed by a more detailed design study, producing a Conceptual Design Report, and a construction project, which could commence by the end of the coming decade. A key input will be the energy scale of whatever new physics is discovered at the LHC; if it is beyond the reach of the ILC, or if (for example) the data reveal a supersymmetric Higgs boson, a muon collider may suddenly become very attractive.

\section{Conclusions}
The long quest for high-intensity muon storage rings appears to be nearing a denouement. This should prove exciting in the coming decade, and  bodes well for the future of high-energy physics!

\begin{acknowledgments}
The author thanks his colleagues of the Neutrino Factory and Muon Collider Collaboration, the MICE Collaboration, Muons, Inc., and the Muon Collider Task Force for many enlightening conversations and interactions over many years. Work supported by DOE grant DE-FG02-01ER41159, NSF grant PHY-0758173, and DOE STTR grant (to Muons, Inc.\ and IIT) AC02-ER86145.

\end{acknowledgments}

\bigskip 

\begin{thebibliography}{99}
\bibitem{other-species}
For example, 40\% survival of antiprotons with deceleration from several MeV down to 40\,keV has been demonstrated in simulation studies~\protect\cite{fridge}.
\end{thebibliography}

\end{document}